\documentclass[twocolumn,showpacs,preprintnumbers]{revtex4}
\usepackage{amsmath}
\usepackage{amssymb}
\usepackage{graphicx}
\usepackage{dcolumn}
\usepackage{bm}
\usepackage{amssymb}
\usepackage{graphicx}
\usepackage{dcolumn}
\usepackage{bm}

\begin{document}

\title{Hyperfine state entanglement of spinor BEC and scattering atom}

\author{Zhibing Li$^1$ \footnote{Z.B.Li, stslzb@mail.sysu.edu.cn}, Chengguang Bao$^1$, Wei Zheng$^2$\footnote{W. Zheng, zhengwei8796@gmail.com}}

\affiliation{$^1$State Key Laboratory of Optoelectronic Materials and Technologies,
Guangdong Province Key Laboratory of Display Material and Technology, School of Physics, Sun Yat-Sen University, Guangzhou, 510275, P. R. China}

\affiliation{$^2$T. C. M. Group, Cavendish Laboratory, J. J. Thomson Avenue, Cambridge CB3 0HE, United Kingdom}

\pacs{}

\begin{abstract}

Condensate of spin-1 atoms frozen in a unique spatial mode may possess large internal degrees of freedom. The scattering amplitudes of polarized cold atoms scattered by the condensate are obtained with the method of fractional parentage coefficients that treats the spin degrees of freedom rigorously. Channels with scattering cross sections enhanced by square of atom number of the condensate are found. Entanglement between the condensate and the propagating atom can be established by the scattering. The entanglement entropy is analytically obtained for arbitrary initial states. Our results also give hint for the establishment of quantum thermal ensembles in the hyperfine space.
\end{abstract}

\maketitle

Quantum entanglement is the centre issue in the modern many-body physics. The entanglement between different parts of the system plays an important role in the thermalization of an isolated quantum system. \cite{Greiner2016} For instance, new phase of matter such as the topological order may be distinguished by the long-range entanglement. \cite{wen2013} On the other hand, generating and manipulating entanglement is the basis of quantum computation and quantum information process.\cite{sten98,reck94} The remote entanglement between quantum objects is the essential ingredient for quantum networks.\cite{Kim08} In experiment, entanglement between an atom and a Bose-Einstein condensate (BEC) has been realized by the intermediate coupling with photon.\cite{let11}

Spinor Bose-Einstein condensates (SBEC) as giant quantum objects which possess large internal degrees of freedom while the spatial modes are frozen have been attracted many interests for testing fundamental concepts of quantum mechanics, for implementation of qubit, and so on. \cite{stam98,kawa12}  Spin mixing experiments have shown that spin states of SBECs may remain coherent from microsecond to second which should be long enough for quantum mechanical manipulation. SBECs are such systems for which feasible method for preparing ground state may exist, but effective method for exact preparation and detecting of a hyperfine spin state is still lacking because the energy splitting of various hyperfine spin states is extremely small due to the tiny spin-spin interaction. \cite{schm04} A possible solution would be to entangle the SBEC with a micro-system that is separated from the SBEC in distance and than to select the wanted SBEC state by measuring the micro-system. In the state collapse of measurement, the entropy of SBEC is transferred to the miro-system (or environment). Consequently, the SBEC has more specific states. The entangled state of a condensate which may consist of billion atoms and a single propagating atom is a perfect analog of the Schr$\ddot{o}$dinger cat state. It would be interesting to implement the which-way experiment \cite{mitt87} and Wheeler's delayed-choice experiment with SBEC-atom scattering.\cite{jacq07,kofl13} One simple way to entangle the BEC with a single atom is using the direct scattering process without any intermediate coupling. Though the entanglement production during the scattering of two particles has been studied, \cite{Schulman2004,Mishima2004,Tal2005,Wang2005,Wang2006,Busshardt2007,Law2004,Harshman2008,Singh2008,Benedict2012} due to the large number of atom and the internal degree of freedom, a SBEC scattered by an atom will give more richer physics.

The present paper studies the entanglement generated by scattering of SBEC and an atom of the same species. It is expected that the spin degrees of freedom of the condensate and the propagating atom will be entangled due to the spin-spin interaction. The spin conservation is also crucial for the entanglement. Therefore, one should go beyond the mean-field theory and deal with the spin degrees of freedom exactly.
We calculate the scattering amplitudes for both elastic and inelastic channels in the single mode approximation (SMA) for the spatial degrees of freedom of the condensate but treating the spin states exactly via the method of fractional parentage coefficients. The entanglement entropy describing quantum entanglement between the scattered atom and the condensate is obtained analytically.

The following discussions will be confined in the case that the incident atom has very low kinetic energy which can not excite the spatial mode of the condensate. Thus the condensate remaining in the spatial ground mode has only spin degrees of freedom. The minimum energy to excite the spatial mode of condensate is about $2\hbar \omega$ with $\omega$ the trapping frequency. As an example for the target condensate of $^{87}$Rb, the incident beam could be a cold atom flux evaporated from another condensate of $^{87}$Rb. If the trapping potential of the target condensate has $\omega=1500$Hz, in order not to excite the spatial mode of the target condensate, each atom of the cold atom flux should have kinetic energy less than $2\hbar\omega\sim peV$, which requires the temperature of the source condensate less than $20n$Kelvin.

Let  $\hat{c}_{\mu}^{+}$ and $\hat{c}_{\mu}$ respectively be the creation and annihilation operators of the spatial ground mode with the spin component $\mu$.
The condensate states of $N$ atoms in the spatial ground mode with total spin $s$ and the magnetization $m$ can be written as $|s,m\rangle=[(\hat{c}^{+})^{N}]_{s,m}|0\rangle$ where $[\cdots]_{s,m}$ means to couple the $N$ spinors to the normalized states that have total spin $s=N, N-2, N-4, \cdots, 1(0)$ if $N$ is odd(even) and the magnetization $m=0, \pm1,\pm2,\cdots, \pm s$. For a given pair of $s$ and $m$, the state $|s,m\rangle$ is unique as proved in \cite{bao04}.
In the low energy scattering, only the s-wave component of the propagating mode has important contribution.  The atom propagating in the s-wave is specified by the wave number $k$. Therefore, the bases of the hyperfine space of the composite system are $\{|k\rangle |w\rangle$, where $|w\rangle=|\mu;s,m\rangle$ with $\mu$ the spin component of the propagating atom.
It has energy $\epsilon_{k}+E_{s}$ with $\epsilon_{k}$ the kinetic energy of the propagating mode and $E_{s}$ the energy of the condensate which only depends on the total spin of the condensate.

Denote the creation and annihilation operators of atom in the s-wave state with wave number $k$ and the spin component $\mu$ as $\hat{b}_{k\mu}^{+}$ and $\hat{b}_{k\mu}$ respectively.
Applying the effective Hamiltonian for the s-wave scattering,\cite{ho98,ohmi98} the interaction responsible to the BEC-atom scattering reads \begin{equation}
 H_{sc}=2\sum_{F=0,2}g_{F}\int_{ k_{1}, k_{2}}\hat{c}_{\mu_{1}}^{+}\hat{b}_{ k_{1}\mu_{2}}^{+}P^{(F)}_{\mu_{1}\mu_{2},\nu_{1} \nu_{2}}\hat{b}_{ k_{2}\nu_{2}}\hat{c}_{\nu_{1}}
 \label{Hsc}
\end{equation}
where $P^{(F)}_{\mu_{1}\mu_{2},\nu_{1}\nu_{2}}=\sum_{M=-F,F}C^{F,M}_{1,\mu_{1},1,\mu_{2}}C^{F,M}_{1\nu_{1},1\nu_{2}}$ with $C^{F,M}_{1,\mu;1,\mu^{\prime}}$ the Clebsch-Gordan coefficients (CGs). The couplings $g_{F}=\frac{4\pi \hbar^{2}a_{F}}{M_{0}}$ with $a_{F}$ the  scattering lengths. \cite{kawa12}
The overall factor $2$ is due to scattering of identical particles.

Consider a polarized initial state $|k_{i}\rangle|w_{i}\rangle$ with $|w_{i}\rangle=|\mu_{i}; s_{i},m_{i}\rangle$ the spin state.
In the distorted wave Born approximation, the spatial mode of the condensate is determined by the trapping potential and the self-interaction of the condensate. Since the incident atom has spin one, the final spin of the condensate can be $s_{\sigma}=s_{i}+2\sigma$ for $\sigma=0,\pm 1$.
Therefore there are nine possible scattering channels, corresponding to the final spin states $|w_{\sigma\mu}\rangle=|\mu ; s_{\sigma},m\rangle$ for $\sigma=0,\pm 1$ and $\mu=0,\pm1$, and $m=m_{i}+\mu_{i}-\mu$ due to the conservation of spin component. The outgoing s-wave of the $\sigma$-channel has the wave number $k_{\sigma}$ which is determined by the energy conservation $\epsilon_{k_{\sigma}}+E_{s_{\sigma}}=\epsilon_{k_{i}}+E_{s_{i}}$.  The channel of $s_{0}=s_{i}$ and $\mu=\mu_{i}$ is the elastic scattering channel. The other eight channels are inelastic channels.

The transition matrix is given by
\begin{eqnarray}
&&\langle w_{\sigma\mu}| \langle k_{\sigma}|H_{sc}|k_{i}\rangle |w_{i}\rangle  \label{trans} \\
&=& 2\sum_{F=0,2}g_{F}P^{(F)}_{\mu_{1}\mu,\nu_{1}\mu_{i}} \langle 0|[c^{N}]_{s_{\sigma} m}c^{+}_{\mu_{1}}
c_{\nu_{1}}[(c^{+})^{N}]_{s_{i}m_{i}}|0\rangle \nonumber
\end{eqnarray}
The annihilation operator $c_{\nu_{1}}$ should contract with each creation operator $c^{+}$ in the factor $[(c^{+})^{N}]_{s_{i}m_{i}}$, which results $N$ identical terms. Thus one only needs to contract $\hat{c}_{\nu_{1}}$ with a specified $c^{+}$ on its r.h.s., the latter should be extracted from  the normalized operator $[(c^{+})^{N}]_{s_{i}m_{i}}$. This can be done with the method of fractional parentage coefficients, which enables to rewrite the normalized operator as
\begin{eqnarray}
&&[(c^{+})^{N}]_{sm}\nonumber \\
&=&\frac{a_{s}^{\{N\}}}{{\sqrt N}}\sum_{\mu} C^{s,m}_{s+1,m-\mu;1,\mu}[(c^{+})^{N-1}]_{s+1,m-\mu}c^{+}_{\mu}\nonumber \\
&~& +\frac{b_{s}^{\{N\}}}{{\sqrt N}}\sum_{\mu}C^{s,m}_{s-1,m-\mu;1,\mu}[(c^{+})^{N-1}]_{s-1,m-\mu}c^{+}_{\mu}
\label{FEB2}
\end{eqnarray}
Where the factors $a_{s}^{\{N\}}$ and $b_{s}^{\{N\}}$ are the fractional parentage coefficients (FPCs) known as \cite{bao05,bao06}
\begin{equation}
a_{s}^{\{N\}}=\sqrt{ \frac{(1+(-1)^{N-s})(N-s)(s+1)}{2N(2s+1)}}       \label{aNSexpr}
\end{equation}
\begin{equation}
b_{s}^{\{N\}}=\sqrt{\frac{(1+(-1)^{N-s})\;s\;(N+s+1)}{2N(2s+1)}} \label{bNSexpr}
\end{equation}
They are nonzero only if $N-S$ is even and the sum of squares of them is one when $N-S$ is even.

The calculation of contractions involving $c^{+}_{\mu_{1}}$ is similar.
All contractions contribute a factor $N^{2}$ to the transition matrix. Both the initial and final states of the condensate have an extra normalization factor $\frac{1}{{\sqrt N}}$ as seen in (\ref{FEB2}). Therefore the transition matrix is enhanced by an overall factor $N$ due to the coherent scattering of $N$ atoms. For large $N$ the condensate scattering will be dominative and the scattering due to the trapping potential is relatively not important and will be neglected.

With the transition matrix calculated  with the FPCs, we obtain the scattering amplitudes $f_{\sigma,\mu}^{i}$ for the channels $|w_{\sigma \mu}\rangle$ as
\begin{eqnarray}
\label{ampl0}
f_{0,\mu}^{i}&=&-4N\sum_{\mu_{1},\nu_{1};F}a_{F}P^{(F)}_{\mu_{1}\mu,\nu_{1}\mu_{i}}\delta_{m-\mu_{1},m_{i}-\nu_{1}} \\
&&\cdot [(a_{s_{i}}^{\{N\}})^{2}C^{s_{i},m}_{s_{i}+1,m-\mu_{1};1\mu_{1}}C^{s_{i},m_{i}}_{s_{i}+1,m_{i}-\nu_{1};1\nu_{1}}
 \nonumber \\
&& + (b_{s_{i}}^{\{N\}})^{2}C^{s_{i},m}_{s_{i}-1,m-\mu_{1};1\mu_{1}}C^{s_{i},m_{i}}_{s_{i}-1,m_{i}-\nu_{1};1\nu_{1}}]\nonumber
\end{eqnarray}
\begin{eqnarray}
f_{\pm,\mu}^{i}&=& -4N\sum_{\mu_{1},\nu_{1};F}a_{F}P^{(F)}_{\mu_{1}\mu,\nu_{1}\mu_{i}}\delta_{m-\mu_{1},m_{i}-\nu_{1}} \\
&~&\cdot b_{s_{i}+1\pm 1}^{\{N\}}a_{s_{i}-1\pm 1}^{\{N\}}C^{s_{i}\pm 2,m}_{s_{i}\pm 1,m-\mu_{1};1\mu_{1}}C^{s_{i},m_{i}}_{s_{i}\pm 1,m_{i}-\nu_{1};1\nu_{1}} \nonumber
\label{amplp}
\end{eqnarray}
The cross sections are proportional to the absolute square of the scattering amplitudes. The branch ratios are given by
\begin{equation}
A^{i}_{\sigma\mu}=\frac{(f^{i}_{\sigma,\mu})^{2}}{\sum_{\sigma,\mu}(f^{i}_{\sigma,\mu})^{2}}
\label{norm}
\end{equation}

The large $N$ behaviors of the amplitudes can be obtained from the CGs and FPs that have explicit expressions. Firstly, the amplitudes of elastic scatterings are always proportional to $N$ as a consequence of coherent scattering.  When $N-s_{i}$ has the same order of $N$, the values of CGs and FPs that fulfil the selection rules of the spin symmetry are of order one, therefore the inelastic amplitudes are also proportional to $N$. From (\ref{aNSexpr}) it can be seen that $a_{s}^{\{N\}}\sim \frac{1}{\sqrt{N}}$ when $s_{i} \sim N$ thus the associate terms can be neglected. If both $s_{i}$ and $|m_{i}|$ are of order $N$ all inelastic amplitudes could be neglected as their scattering branch ratios with respect to the elastic amplitudes are vanishing as $\frac{1}{\sqrt{N}}$ or even faster. When $s_{i}$ is of order $N$ but $|m_{i}|$ of order one, the inelastic amplitudes of the $\sigma=0$ channel is also proportional to $N$. Four examples are given in the followings (only channels with amplitudes proportional to $N$ are retained and the difference between $a_{0}$ and $a_{2}$ has been neglected).

(i) The initial states with $s_{i}=m_{i}=N$ and $\mu_{i}=0,-1$ has only elastic scattering amplitude proportional to $N$.
Amplitudes of all inelastic channels are at most proportional to $\sqrt{N}$ therefore have negligible contribution.

(ii) For $s_{i}=N$ and $m_{i}=\mu_{i}=0$, there are three possible final spin states:
$|0;N,0\rangle$ and $|\pm 1;N, \mp 1\rangle$. The first one is from the elastic scattering, having normalized amplitude  $\frac{3}{\sqrt{10}}$. The other two have $\frac{1}{2\sqrt{5}}$.

(iii) For  $s_{i}=m_{i}=0$ and $\mu_{i}=0$, the elastic scattering final spin state is
$|0;0,0\rangle$, having normalized amplitude  $\frac{2\sqrt{2}}{3}$; while the inelastic scattering final spin states are  $|0;2,0\rangle$ and $|\pm 1;2, \mp 1\rangle$  with  normalized amplitudes $-\frac{1}{3}\sqrt{\frac{2}{5}}$, and $\frac{1}{\sqrt{30}}$ respectively.

(iv) For  $s_{i}=m_{i}=0$ and $\mu_{i}=-1$, the elastic scattering final spin state is
$|-1;0,0\rangle$, having normalized amplitude  $\frac{2\sqrt{2}}{3}$; while the inelastic scattering final spin states are  $|-1;2,0\rangle$, $|0;2,-1\rangle$ and $|1;2,-2\rangle$, having normalized amplitudes $\frac{1}{3\sqrt{10}}$,  $-\frac{1}{\sqrt{30}}$, and $\frac{1}{\sqrt{15}}$, respectively.

The scattering is dominated by the elastic channel having over $85\%$ branch ratio for any initial state. Since the elastic scattering does not change the state of the condensate, it is less interesting. To exclude the elastic scattering, one can use a spin filter, such as Stern-Gerlach apparatus, that separates the outgoing beam into three ones according to their spin polarization and absorbs the beam of the same spin polarization as the incident flux. The left two beams will all come from inelastic scattering. Coherent superpositions of the left two spin-flipped beams can be recovered by another Stern-Gerlach apparatus that combines two beams and erases the information of paths which any beam would have taken. The effect of the spin filter is just decreasing the brightness of the incident flux. We can subtract the elastic scattering wave from the entire outgoing wave and concentrate on the spin-flipped scattering. On that account, we introduce a spin-flipped branch ratio (SFBR) as the cross section ratio of a spin-flipped channel over all of those. When the elastic scattering background is subtracted, the scattering amplitudes depend on $a_{F}$ significantly. In the following we will consider the condensate of $^{87}$Rb that has $a_{0}=101.8$ and $a_{2}=100.4$ (a.u.).

First, consider the incident flux with $\mu_{i}=-1$. The SFBR of $\mu=0,1$ versus $\frac{m_{i}}{N}$ for $s_{i}=\frac{N}{2},N$ is given in Fig.1. It is clear that the final condensate magnetization $m\le m_{i}$ because the conservation of spin component associate with $\mu_{i}=-1$. For $s_{i}=N$, the amplitude of $\mu=0$ ($\mu=1$) is a monotonic decreasing (increasing) function of $m_{i}$. But generally the amplitudes have complicate behavior.
To compare the transition probabilities of spin increased and spin decreased channels, the differences of the SFBRs of the $\sigma=\pm$ channels is plotted in Fig.2. The triangle is the hyperfine space of the condensate where the left apex has $s=m=0$ and the vertical line on the right has $s=N$ and $m$ running from $-N$ to $N$ from the lower end to the upper end. The condensate spin tends to increase in the positive regime and tends to decrease in the negative regime. Combining the information of Fig.1 and Fig.2, one can have the following picture of state evolution of condensate in steady scattering of $\mu_{i}=-1$ incident flux: the states in the upper-right regime flow to the left and approach the horizontal axis of $m=0$; after crossing the axis to the regime of negative $m$ the states will flow to the lower-right apex and finally the condensate maximizes its magnetization to $-N$. If the incident flux has $\mu_{i}=1$ the condensate will reversely go to the state of $s=m=N$.

Second, consider the scattering of the zero-polarized incident flux with $\mu_{i}=0$.
The SFBR of $\mu=\pm 1$ is given in Fig.3 for $s_{i}=\frac{N}{2},N$. The amplitude of the $\mu=1$ is larger (smaller) than that of the $\mu=-1$ for $m_{i}>0$ ($m_{i}<0$). Therefore the magnetization of the condensate tends to vanish in the scattering of the zero-polarized incident flux. Fig.4 shows the SFBR for the initial magnetization $m_{i}=0$ and $\frac{N}{2}$. The SFBRs of $\sigma=\pm $ channels have observable difference only in the small regime near to $s_{i}=0$.  In the steady scattering, the states will approach the $m=0$ axis from both sides in the hyperfine space. The condensate tends to increase its total spin in the large spin regime but very slowly.

The reduced density of matrix of the condensate after scattering is given by tracing off the outgoing states,
\begin{equation}
\rho_{c}=\sum_{\sigma,\mu}A^{i}_{\sigma\mu}|s_{\sigma},m\rangle \langle s_{\sigma},m|
\label{rmcond1}
\end{equation}
where $A^{i}_{\sigma\mu}$ is the branch ratio of the $(\sigma,\mu)$ channel. If the elastic scattering channel is excluded, $A^{i}_{\sigma\mu}$  should be the SFBR.
Correspondingly, the entanglement entropy of the system is given by
\begin{equation}
S_{i}=-\sum_{\sigma,\mu} A^{i}_{\sigma\mu} \log ( A^{i}_{\sigma\mu} )
\label{entr1}
\end{equation}

The final state of the case (i) discussed previously has $S_{i}=0$ therefore it is not an entangled state. The case (ii) has entanglement entropy $0.3944$ when all channels, including the elastic channel, are considered. If the elastic channel is excluded, the entanglement entropy of the case (ii) becomes $\ln(2)$. Excluding the corresponding elastic channels, the case (iii) and (iv) have entanglement entropies $1.0889$ and $0.8979$ respectively. The dependence of the entanglement entropy on the initial states, with the elastic component subtracted, is presented in Fig.5: for $\mu_{i}=0$ (left) and for $\mu_{i}=-1$ (right). The case of $\mu_{i}=1$ can be obtained from Fig.5(right) as a reflection with respect to the horizontal axis.

Generally $S_{i}$ is nonzero, implying that the outgoing atom and the condensate is entangled. A deterministic measurement of the polarization of the outgoing beam will cause the state of the condensate collapse. Whenever such outgoing state is measured or operated quantum mechanically, the condensate will collapse to a specific state instantly. This suggests a method for superfine states preparation. On the other hand, the condensate will be described by the non-pure state with probability for each spin state given by (\ref{rmcond1}) if the information of outgoing atoms is completely lost.

In summary, we have obtained the scattering amplitudes of condensate-atom scattering, dealing the spin degrees of freedom strictly by the method of fractional parentage coefficients. It is found that the elastic scattering channel as well as some inelastic scattering channels are enhanced by square of atom number of the condensate.  Due to this notable fact the channels would be observable. The scattering branch ratios imply that the zero-polarized incident flux vanishes the magnetization of the condensate, while the nonzero-polarized incident flux maximizes the magnetization. In steady scattering, the flow of states in the hyperfine space of the condensate is outlined. The flow is mainly determined by the spin symmetry and weakly if any depends on the scattering lengths. It provides a method to create the occupation reversed states in the hyperfine space. For some special cases, such as the cases (ii) to (iv) discussed in the text, the quantum collapse of the outgoing atom can specify a hyperfine state of the condensate. The analytical expression for the entanglement entropy as a function of the initial states is given in this letter. Except some special initial states, the entanglement entropy is generally nonzero, implying that the scattered atom and the condensate is entangled. The proposed process could be implemented at nano-Kelvin theoretically and would be interesting for the studies of quantum information as it provides a new way to manipulate the degenerate quantum states of gigantic systems that would be useful in quantum information.

\begin{acknowledgments}
The project is supported by the National Natural Science Foundation of China (Grant:
11274393), the National Basic Research
Program of China (Grant: 2013CB933601), and the National Key Research and Development Project of China
(Grant: 2016YFA0202001).
\end{acknowledgments}

\clearpage

\begin{figure}[tbp]
\includegraphics{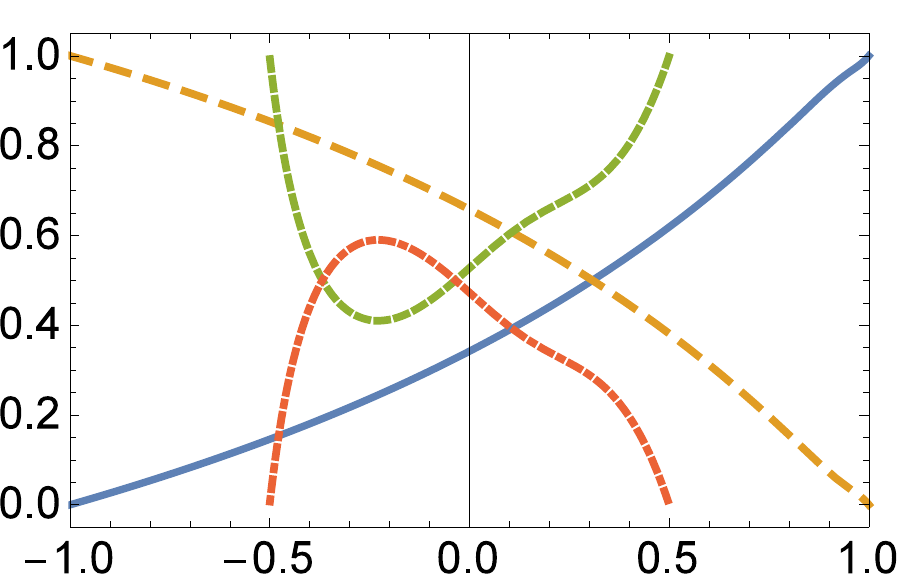}
\caption{(color online) The spin-flipped branch ratios among the inelastic channels of $\mu=0$ and $1$ for the initial incident polarization $\mu_{i}=-1$. The horizontal axis is $\frac{m_{i}}{N}$. The solid and dashed curves are SFBRs of $\mu=0$ (solid) and $\mu=1$ (dashed) for the initial condensate spin $s_{i}=N$.  The short-dashed and dash-dotted curves respectively are SFBRs corresponding to $\mu=0$ and $\mu=1$ for $s_{i}=\frac{N}{2}$.}
\label{fig1}
\end{figure}

\begin{figure}[tbp]
\includegraphics{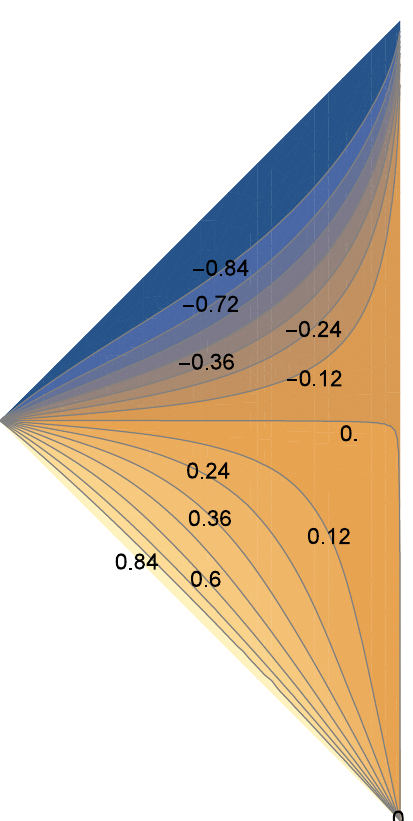}
\caption{The contour plot of the difference of the branch ratios between the $\sigma=1$ and $-1$ channels for the initial incident polarization $\mu_{i}=-1$. The values labelled the contour lines are the differences. The initial condensate spin $s_{i}$ runs from $0$ to $N$ along the horizontal line from the left point to the right edge of the triangle. The initial condensate magnetization $m_{i}$ runs from $-s_{i}$ to $s_{i}$ along a vertical line from the lower edge to the higher edge of the triangle.}
\label{fig2}
\end{figure}

\begin{figure}[tbp]
\includegraphics{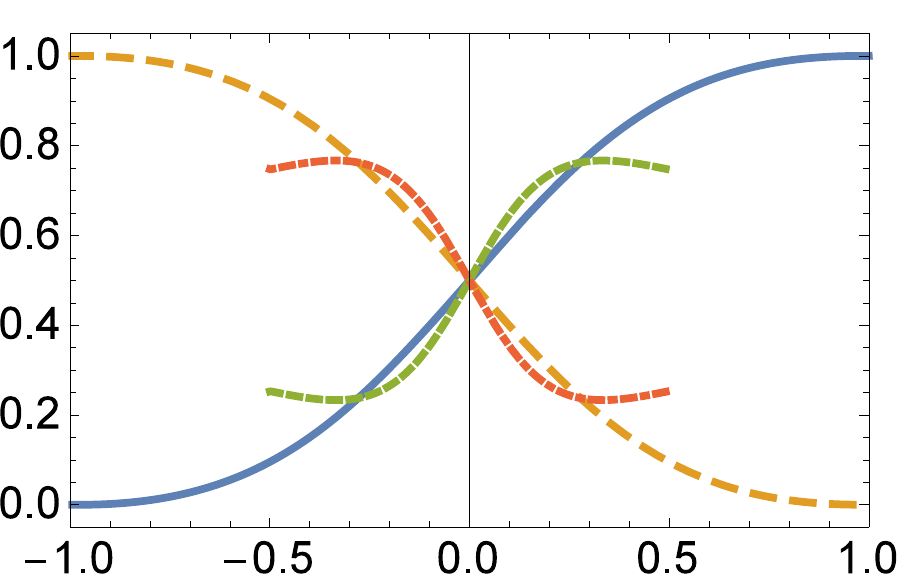}
\caption{(color online) The spin-flipped branch ratios of $\mu=\pm 1$  for the initial incident polarization of $\mu_{i}=0$. The horizontal axis is $\frac{m_{i}}{N}$. The solid and dashed curves are SFBRs of $\mu=1$ (solid) and $\mu=-1$ (dashed) for the initial condensate spin $s_{i}=N$.  The short-dashed and dash-dotted curves respectively are SFBRs corresponding to $\mu=1$  and $\mu=-1$ for $s_{i}=\frac{N}{2}$.}
\label{fig3}
\end{figure}

\begin{figure}[tbp]
\includegraphics{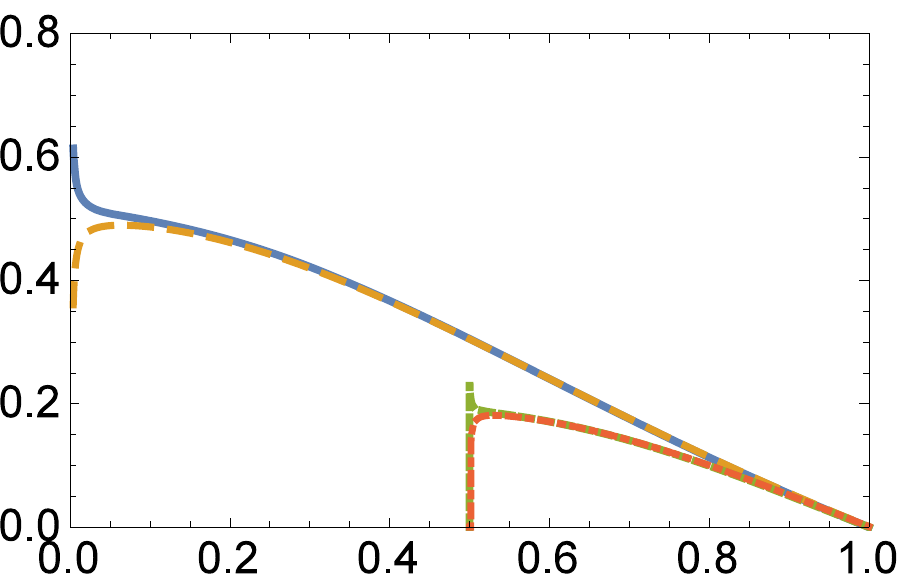}
\caption{(color online) The spin-flipped branch ratios among the inelastic channels of $\sigma=\pm 1$ for the initial incident polarization of $\mu_{i}=0$. The horizontal axis is $\frac{s_{i}}{N}$. The solid and dashed curves are SFBRs of $\sigma=1$ (solid) and $\sigma=-1$ (dashed) for the initial condensate magnetization of $m_{i}=0$.  The short-dashed and dash-dotted curves respectively are SFBRs of $\sigma=1$ and $\sigma=-1$ for $m_{i}=\frac{N}{2}$. }
\label{fig4}
\end{figure}

\begin{figure}[tbp]
\includegraphics{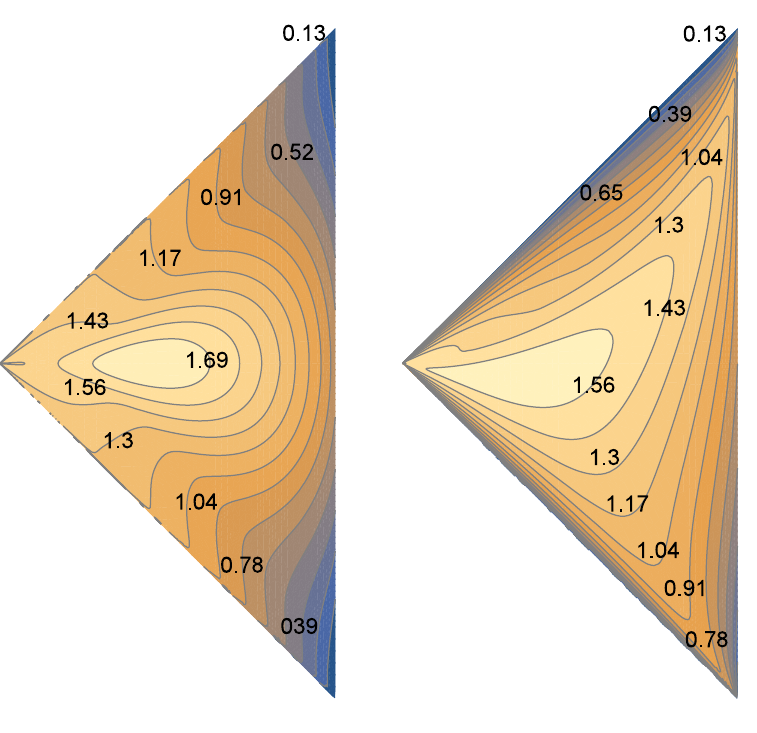}
\caption{The entanglement entropy for the initial incident polarizations $\mu_{i}=0$ (left) and  $\mu_{i}=-1$ (right). The values labelled the contour lines are the entanglement entropy of the final states with the elastic component subtracted. The triangles have the same meaning as Fig.2.}
\label{fig5}
\end{figure}

\end{document}